\documentclass[%
 reprint,nofootinbib,
superscriptaddress,
 amsmath,amssymb,
 aps,
floatfix,
]{revtex4-2}
\usepackage{braket}
\usepackage{color}
\usepackage{graphicx}% Include figure files
\usepackage{dcolumn}% Align table columns on decimal point
\usepackage{bm}% bold math
\usepackage[
	  pagebackref=false,
	  colorlinks=true,
      linkcolor=blue,
      urlcolor=blue,
      filecolor=black,
      citecolor=blue,
      pdfstartview=FitV,
      pdftitle={},
        pdfauthor={},
        pdfsubject={},
        pdfkeywords={},
        pdfpagemode=None,
        bookmarksopen=true
      ]{hyperref}

\begin{document}

\preprint{APS/123-QED}

\title{Operator and entanglement growth in non-thermalizing systems: many-body localization and the random singlet phase}

\author{Ian MacCormack}%
 \email{imaccormack@uchicago.edu}
\affiliation{Kadanoff Center for Theoretical Physics, University of Chicago, Chicago, IL~60637, USA}
\affiliation{James Franck Institute, University of Chicago, Chicago, Illinois 60637, USA}
 \author{Mao Tian Tan}
\email{mtan1@uchicago.edu}
\affiliation{Kadanoff Center for Theoretical Physics, University of Chicago, Chicago, IL~60637, USA}
\affiliation{James Franck Institute, University of Chicago, Chicago, Illinois 60637, USA}
 \author{Jonah Kudler-Flam}
\email{jkudlerflam@uchicago.edu}
\affiliation{Kadanoff Center for Theoretical Physics, University of Chicago, Chicago, IL~60637, USA}
\affiliation{James Franck Institute, University of Chicago, Chicago, Illinois 60637, USA}
\author{Shinsei Ryu}%
 \email{shinseir@princeton.edu}
\affiliation{
Department of Physics, Princeton University, Princeton, New Jersey 08540, USA
}

\date{\today}

\begin{abstract}
We characterize the growth and spreading of operators and entanglement in two paradigmatic non-thermalizing phases -- the many-body localized (MBL) phase and the random singlet phase (RSP) -- using the entanglement contour and multipartite operator entanglement measures. The entanglement contour characterizes the spacetime spreading of entanglement. We find a logarithmically growing \textit{entanglement light cone} in the MBL phase and a power-law entanglement light cone for the RSP, sharply contrasting the linear light cones of clean, thermalizing systems. The operator entanglement characterizes \textit{scrambling} i.e.~the delocalization of information. We find slow scrambling behavior in the MBL phase; the late-time value of the tripartite mutual information scales linearly with system size, but is subthermal. In contrast, the tripartite logarithmic negativity scales sublinearly, revealing an intriguing distinction between classical and quantum information scrambling in the MBL phase. In the RSP, there is no signature of scrambling. Contrasting with out-of-time-ordered correlators, we emphasize the different velocities at which state and operator information propagates.
\end{abstract}

\maketitle

% \tableofcontents
\section{Introduction}

% In recent years, the study of chaos and thermalization in out-of-equilibrium
% quantum systems has received much attention.
Generically, interacting quantum systems pushed out of equilibrium with some
finite energy density rapidly reach equilibrium and become describable (locally) by a thermal ensemble.
This can be understood as a process of information loss, in which the local details of the initial state are spread out, or scrambled, across all degrees of freedom, becoming inaccessible to local measurements. 

While most works on quantum chaos and thermalization have focused on systems that obey the eigenstate thermalization hypothesis (ETH) \cite{PhysRevA.43.2046,1994PhRvE..50..888S} and rapidly scramble information, recent studies of disordered systems have found signatures of \textit{slow scrambling} in out-of-time-ordered correlators (OTOC) \cite{2016arXiv160801914F,2016arXiv160802765C, 2017AnP...52900332C,2017PhRvB..95f0201S, 2017AnP...52900318H,2017PhRvB..95f0201S, 2018PhRvB..98c5118N, 2018PhRvX...8c1057K, 2018arXiv180706086S, 2019arXiv190207199X}, revealing rich dynamical structure in systems that do not reach thermal equilibrium, or do so exponentially slowly. In order to better understand slowly scrambling behavior, we analyze the many-body localized (MBL) phase and the random singlet phase (RSP) in two distinct steps: (i) We compute the entanglement contour, which serves as a well-behaved entanglement density function \cite{2014JSMTE..10..011C}, following a global quench to characterize entanglement spreading in spacetime. This reveals emergent \textit{entanglement light cones} distinct from the light-cones found in OTOCs. (ii) We compute the tripartite mutual information of the time-evolution operator. This characterizes the extent to which initially localized information delocalizes (scrambles) across the entire system. Two systems with the same entanglement light cone structure may have different behaviors of the operator entanglement. The operator entanglement being nontrivial implies that entanglement is being produced, complexified, and delocalized rather than simply spread around the system coherently. We now review the definitions of these two main actors.

\paragraph*{Entanglement Contour}
The entanglement contour has been shown to provide an intuitive picture of the extent to which each degree of freedom in a given subsystem is entangled with the subsystem's complement. A natural proposal\footnote{Though not uniquely defined, this definition has been shown to give nearly identical results as other proposals specific to free systems \cite{2020ScPP....8...63K}. Moreover, it has been argued to be unique once imposing an additional physical requirement \cite{2020PhRvR...2b3170W}} for generic many-body systems is to partition the subsystem, $A$, into subsets $\{ A_i\}$
\cite{2018arXiv180305552W,2019arXiv190204654K,2019arXiv190206905W,2020PhRvR...2b3170W} 
\begin{align}
    s_A(A_i)\equiv \frac{1}{2} [ S(A_i &| A_1 \cup \dots \cup A_{i-1}) 
    \nonumber
    \\
    &+ S(A_i | A_{i+1}\cup \dots \cup A_n)  ],
    \label{contour_formula}
\end{align}
where $S(A|B)$ is the conditional entropy
\begin{align}
    S(A|B) \equiv S(A\cup B)-S(B),
\end{align}
and $S(A)$ is the von-Neumann entanglement entropy associated to (the reduced
density matrix of) the subsystem $A$.
When we take the continuum limit or consider the contour on a single lattice
site, we often use the notation $s_A(x)$ where $x$ denotes the spatial position
of $A_i$.
The entanglement contour can be viewed as a well-behaved entanglement density
function:
$\sum_i s_A(A_i) = S(A)$
or
$\int_A dx\, s_A(x) = S(A)$.
In other words, the entanglement contour can be viewed as a spatially resolved
version of the entanglement entropy $S(A)$,
and, when applied to time-dependent states,
provides a quasi-local picture of how entanglement entropy spreads in a system.

%\begin{widetext}
\begin{table*}
\centering
\scriptsize
\begin{tabular}{ | p{.1\textwidth} | p{.28\textwidth}| p{.28\textwidth} | p{.28\textwidth}| } 
\hline
System  & Half-space $S_A (t)$ after Global Quench & Half-space $s_A(x,t)$ after Global Quench&  TOMI/TOLN $(t\rightarrow \infty)$\\ 
\hline
Free Fermion CFT& 
$
\frac{1}{12} \log \left(\frac{\beta}{\pi \epsilon } \sinh\left(\frac{2\pi t}{\beta} \right)\right)
$
taking the thermodynamic limit, otherwise, there will be revivals \cite{2005JSMTE..04..010C, 2014PhRvL.112v0401C}. 
%$\epsilon$ is the UV cutoff and $\beta$ is the effective temperature of the quench.
& 
$
\begin{cases}
0 & |x| < 2t \\
\frac{c \pi}{12\beta}\coth\left(\frac{\pi x}{\beta}\right)  & |x| > 2t
\end{cases}
$ 
\cite{2014JSMTE..10..011C,2019arXiv190204654K, 2019arXiv190501144D}. This describes a linear light cone.
& $I_3 = \mathcal{E}_3 = 0$  \cite{2019JSMTE..09.3107N,2019arXiv190607639K} i.e.~no scrambling.\\ 
\hline
Holographic CFT &
$
\frac{c}{12} \log \left(\frac{\beta}{\pi \epsilon } \sinh\left(\frac{2\pi t}{\beta} \right)\right)
$
taking the thermodynamic limit, otherwise, there will be revivals \cite{2013JHEP...05..014H, 2016arXiv160407830M}.  %$c$ is the central charge.
& 
$
\begin{cases}
0 & |x| < 2t \\
\frac{c \pi}{12\beta}\coth\left(\frac{\pi x}{\beta}\right)  & |x| > 2t
\end{cases}
$ 
\cite{2019arXiv190204654K}. This describes a linear light cone.
& $I_3 = - 2 S_A$ \cite{2019JSMTE..09.3107N} 
$\mathcal{E}_3 = - 2 S_A^{(1/2)}$ \cite{2019arXiv190607639K} These are thermal entropies and are of the largest magnitude possible i.e.~maximal scrambling.\\ 
\hline
RSP & Quasi-power law growth at early times (Fig.~\ref{fig:RSPEEgrowth}) and $\log \big (\log (t) \big )$ at very late times \cite{2012PhRvB..85i4417I}. & A power law light cone emerges of the form $x_c \sim t^\alpha$, with $\alpha \sim 0.264$. There is a $\sim x^{-1}$ decaying profile at long times (Fig.~\ref{fig:rsp_quench}). 
& $I_3$ and $\mathcal{E}_3$ saturate to a constant value, with no length dependence implying no scrambling for large system sizes.\\ 
\hline
MBL & Logarithmic growth and saturation to sub-thermal volume-law in finite systems \cite{2008PhRvB..77f4426Z, 2012PhRvL.109a7202B, 2013PhRvL.110z0601S}. & A logarithmic light cone emerges with an exponentially decaying profile at late times (Fig.~\ref{fig:LLC2}).
& $I_3$ and $\mathcal{E}_3$ both saturate to macroscopic negative values that scale with the size of region $A$ at exponentially long times with the saturation value of $I_3$ scaling linearly with $A$ and $\mathcal{E}_3$ potentially sublinearly. Neither are maximal (Fig.~\ref{fig:mbl_operator}).  \\ 
\hline
Anderson Localization & Rapid saturation to area-law entanglement \cite{2008PhRvB..77f4426Z,2012PhRvL.109a7202B}. & Information is localized i.e.~no spreading.
& $I_3, \mathcal{E}_3 \sim 0$ i.e.~no scrambling.\\ 
\hline
\end{tabular}
\caption{
% Table summarizing what is
Summary of the results of the paper and other known results about
the entanglement contour $s_A(x,t)$
and tripartite operator entanglement (TOMI and TOLN)
in various classes of systems.
The behavior of the entanglement entropy
$S_A(t)$ for a subregion (an interval) $A$
after global quench is also presented.
Here, 
$c$ is the central charge,
$\epsilon$ is the UV cutoff,
and 
$\beta$ is the effective 
temperature of the quench
determined by the energy 
injected to the system. We use the free fermion CFT as an example of a system with nontrivial
entanglement spreading but no scrambling, holographic CFTs as examples of maximal scramblers,
and Anderson localized systems as non-scramblers with no entanglement
spreading.
%  {\color{red}What are $L, \beta, c, \epsilon$ in the table?
%    For $S_A(t)$ the size of the interval does not matter?
%  }\textcolor{blue}{[Defined (or removed) all variables. Clarified that we are considering the half-space entropy in the thermodynamic limit, so there is no interval size dependence. ]}
}
\label{Tab:tab1}
\end{table*}
%\end{widetext}

\paragraph*{Operator Entanglement}
To define the operator entanglement \cite{2001PhRvA..63d0304Z},
% , PhysRevA.76.032316,PhysRevB.79.184416,2016JHEP...02..004H,Dubail_2017,PhysRevLett.122.250603,Xu2019,2019JSMTE..09.3107N,2019arXiv190607639K,Wang2019}
we first map the time evolution operator, $U(t)$, to a state in a doubled Hilbert space, $\mathcal{H}_1 \otimes \bar{\mathcal{H}}_2$, under the channel-state duality \cite{CHOI1975285, JAMIOLKOWSKI1972275}. Explicitly, the time evolution operator may be expanded in its energy eigenbasis as
\begin{align}
    U(t) = e^{-i H t} \sum_i \ket{i}\bra{i}.
\end{align}
We can then dualize 
the bra vector to define the state
\begin{align}
    \ket{U(t)} = \mathcal{N} e^{-iH_1 t}\sum_i \ket{i}_1 \ket{i^{*}}_2,
    \label{tfd}
\end{align}
where we take the CPT conjugate and $\mathcal{N}$ is a normalization factor. The Hamiltonian acts only on the first copy of the Hilbert space. We then compute entanglement measures within this state. 
Throughout this paper, we let $A$ be a subsystem in the ``input'' Hilbert space, $\mathcal{H}_1$, and $B $, $C$ be subsystems in the ``output'' Hilbert space, $\mathcal{H}_2$, with $B\cup C = \mathcal{H}_2$. Using this partitioning, we can compute the bipartite operator mutual information (BOMI) using the standard definition of mutual information in terms of operator entanglement entropies 
\begin{equation}
    I(A,B)= S(A) + S(B) - S(A \cup B),
\end{equation}
and tripartite operator mutual information (TOMI) by a taking linear combination of BOMIs 
\begin{align}
    I_3(A,B,C) &= I(A,B) + I(A,C) - I(A,B\cup C) .
\end{align}
The TOMI is symmetric about the three regions and diagnoses the extent to which information is delocalized in the quantum channel \cite{2016JHEP...02..004H}.

One disadvantage of the mutual information is that it is sensitive to both quantum and classical correlations. To isolate the quantum information that is scrambled, we compute bipartite operator logarithmic negativity (BOLN) i.e.~the logarithmic negativity, $\mathcal{E}(A,B) \equiv \log  ( | \rho_{AB}^{T_B}|_1 )$, in the operator state. In analogy to TOMI, we also study the tripartite operator logarithmic negativity (TOLN)
\begin{equation}
    \mathcal{E}(A: B,C) = \mathcal{E}(A,B) + \mathcal{E}(A,C) - \mathcal{E}(A,B\cup C).
\end{equation}
The TOLN characterizes the amount of purely quantum information scrambled by the quantum channel. In the MBL phase, we find this to have qualitatively different behavior than the TOMI.

\subsection{Summary of Results}

Before proceeding to the details of our analysis, we present Table \ref{Tab:tab1} that summarizes our results and contrasts them with previously known results for representative classes of quantum systems. The upshot is that we find a logarithmic light cone for the MBL phase and a novel power-law light cone in the RSP. These sharply contrast the linear entanglement light cones observed in both free and holographic theories. Moreover, we find that the exponentially late-time values for the TOMI in the MBL phase scale with the size of region $A$ but do not saturate the fundamental bounds that are saturated in e.g.~random unitary circuits and holographic conformal field theories \cite{2019JSMTE..09.3107N, 2019arXiv190607639K}. Though we are restricted to relatively small system sizes, the TOLN appears to scale sublinearly. This means that MBL systems scramble classical and quantum information non-trivially, albeit exponentially slowly. In contrast, we find the RSP to have trivial TOMI/TOLN, meaning that the RSP channel does not scramble information even though it does spread in spacetime. We attribute this to the free fermion realization of the RSP that we use.

\section{Many-Body Localization}
\label{MBL_sec}
Many-body localization is perhaps the best known example of ergodicity-breaking in many-body quantum systems. It occurs in interacting systems when an on-site potential is tuned to be sufficiently spatially disordered. MBL has been a subject of intense study in recent years. See, for example, the recent review Ref.~\cite{2017AnP...52900169A} and references there within. As the strength of the on-site disorder is increased relative to the interaction strength, more and more of the system's high energy eigenstates turn from typical volume-law entanglement states (as the eigenstate thermalization hypothesis would imply), to short-range entangled area-law states \cite{2013PhRvL.111l7201S}. 
Once the localization transition is passed, all eigenstates of the system become area-law entangled, and the system is fully many-body localized. Before reaching the transition, it is possible to have a mobility edge, separating area-law states from volume-law states.
In addition to the area-law eigenstates, MBL systems display a number of interesting features.
In the localized phase, the systems contain an extensive number of emergent local integrals of motion (LIOMs, sometimes called ``l-bits"), with exponentially decaying spatial support. 
One can write an effective Hamiltonian for the MBL system in terms of these LIOMs \cite{2013PhRvL.111l7201S,2014arXiv1408.4297H}. 

MBL systems are also fascinating from a dynamical perspective. Generic interacting many-body quantum systems are thought to be ergodic in the sense that after sufficiently long time evolution, expectation values of local operators become exponentially close to their corresponding thermal expectation values. Thus, memory of the initial state becomes inaccessible to local measurements and any subsystem can be described by a small number of thermodynamic quantities. MBL systems, on the other hand, do not thermalize. The conserved LIOMs serve to retain memory of the initial state, precluding a description of the late-time state by a thermal ensemble. MBL provides an important counterexample to the eigenstate thermalization hypothesis (ETH) and motivates us to further understand when and how closed quantum systems fail to thermalize. We additionally note that MBL phases and their non-ergodic properties have been directly observed in a number of experimental settings \cite{2016NatPh..12..907S, 2015arXiv150901160I, 2019Sci...364..256L}.

Despite their lack of energy and particle transport, MBL systems nevertheless produce nontrivial long-range entanglement in far-from-equilibrium scenarios. For example, if we quench into an MBL Hamiltonian starting from a product state, the entanglement entropy grows logarithmically in time for a time that scales exponentially with the system size \cite{2008PhRvB..77f4426Z, 2012PhRvL.109a7202B}. After this time, the saturated entanglement entropy in finite systems displays a volume-law, though with a smaller multiplicative coefficient than the volume-law for the thermal state \cite{2012PhRvL.109a7202B}. In some sense, this volume-law indicates a partial thermalization of finite size MBL systems \cite{2013PhRvL.111l7201S}. We are motivated to study this novel entanglement growth more deeply. In particular, we would like to understand how the entanglement spreads at a local level and if/how the information becomes delocalized.

To ensure we capture universal features of the MBL phase, and no artifacts of a particular model, we use the phenomenological MBL fixed point Hamiltonian constructed from the local integrals of motion

\begin{equation}
    H_{LIOM} = - \sum_i h_i \sigma^z_i - \sum_{i<j} J^{(2)}_{ij}  \sigma^z_i \sigma^z_j- \sum_{i<j<k} J^{(3)}_{ijk}  \sigma^z_i \sigma^z_j\sigma^z_k + ...
\end{equation}
The interaction terms are defined as

\begin{equation}
    J^{(2)}_{ij} = \frac{1}{2} J_{ij} \exp(-|i-j|/\xi ) \quad J^{(3)}_{ijk} = \frac{1}{6} J_{ijk} \exp(-|i-k|/\xi ),
\end{equation}
where $J_{ij}$ and $J_{ijk}$ are drawn randomly (from a uniform distribution in our case, but other distributions can be used) and $\xi$ is a dimensionless localization length. This model was used to compute the OTOC in \cite{2017AnP...52900332C, 2017PhRvB..95f0201S}. We also draw $h_i$ from the same distribution. The Pauli operators $\sigma^z_i$ denote emergent local integrals of motion, which are exponentially localized in the physical lattice. For our purposes, we treat them as perfectly localized, for the sake of computation.

\subsection{A Logarithmic Entanglement Light Cone}

\begin{figure}
  \centering
  \includegraphics[width=8cm]{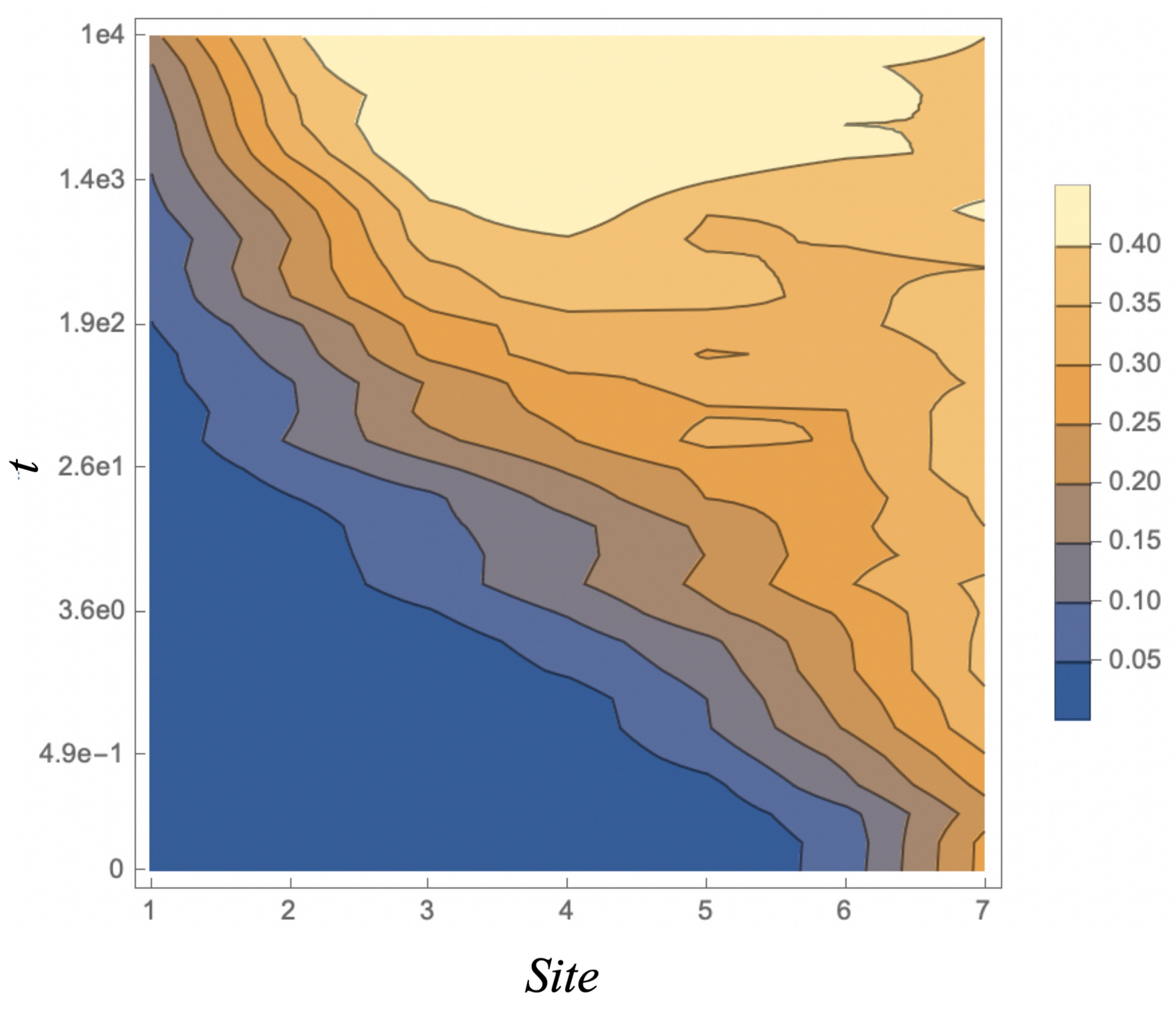}
  \caption{The entanglement contour after a global quench from $H_{LIOM}$ averaged over 40 disorder realizations, normalized by $\log 2$, and smoothed out to remove pixelation. The quench was performed from random product states on $14$ sites using a disorder strength of $J=100$. The contour depicted describes the seven leftmost sites. The level sets make the logarithmic entanglement light cone clear (modulo the edge effects occurring at sites 1 and 2).}
  \label{fig:LLC2}
\end{figure}

We study the entanglement contour after a global quench in order to identify how much entanglement entropy each site contributes to the overall logarithmic growth. Before turning to numerics, we motivate an analytical prediction for the form of the entanglement contour using the language of the emergent LIOMs.

The effective interaction between two ``l-bits" separated by a distance $d$ is given by \cite{2015ARCMP...6...15N}
\begin{equation}
    J^{{\it eff}} \sim J_0 \exp(-d/\xi),
\end{equation}
where $J_0$ is the bare interaction (in our case, $J_0 = J_{zz} = 1$) and $\xi$
is the localization length. Using the effective interaction of the dressed
spins, one can obtain an estimate for the amount of time it takes for two
unentangled l-bits to become entangled. This happens when $J^{{\it eff}} t \geq 1$, so
\begin{equation}
    t \sim \frac{1}{J_0} \exp ( -d/\xi).
\end{equation}
We now consider an MBL system on a chain of total length $L$, with a subinterval $\lbrack0,\ell \rbrack$. Picking a point $x<\ell$ within this subinterval, we can count the number of l-bits outside of the interval with which the l-bit at point $x$ is entangled at a particular time $t$. This is precisely what the entanglement contour should describe. The result (up to proportionality constants) is
\begin{gather}
    s_{A} (x,t) \propto \begin{cases} 
      \displaystyle
    0 & t <  \frac{
    e^{  \frac{\ell -x}{\xi} }
    }{J_0} 
    \\ \\
      \displaystyle
    \frac{1}{L} \big (\xi \log (J_0 t) +x -\ell \big) & 
    \frac{
    e^{ \frac{\ell -x}{\xi} } 
    }{J_0} 
    \leq  t 
     \leq  \frac{
    e^{  \frac{L -x}{\xi} }
    }{J_0} 
    \\ \\
      \displaystyle
    1 - \frac{\ell}{L} & t >  \frac{
     e^{ \frac{L -x}{\xi} }
    }{J_0}
    \end{cases}
    \label{mbl_contour_estimate}
\end{gather}

The form of a logarithmic light cone is clear; the wave front of the contour arrives at a time exponential in the distance from the entangling surface $t = \frac{1}{J_0} \exp \left( \frac{\ell -x}{\xi} \right) $. 
Once this time has passed, the magnitude of the contour increases linearly, saturating at a constant value. This agrees with the observation that entanglement entropy grows logarithmically with time in MBL, eventually reaching a volume-law in the long time limit. Let us now turn to numerics in order to verify \eqref{mbl_contour_estimate}.

We select a subinterval consisting of the leftmost seven sites of our fourteen site chain and compute the entanglement contour of this interval after the global quench using $H_{LIOM}$. 
We observe a distinct logarithmic light cone in the entanglement contour in Fig.~\ref{fig:LLC2}. A similar logarithmic light cone has previously been observed in out-of-time-order correlators (OTOCs) of certain local operators in the MBL phase \cite{2017AnP...52900318H}. These are related but distinct light cones. 

The authors of Ref.~\cite{2017AnP...52900318H} computed the OTOC as a function of space and time in the disordered Heisenberg model.
Defining the butterfly velocity, $v_B$, as
\begin{equation}
    j_{\epsilon} \sim v_B \log_{10} t,
\end{equation}
where $j_{\epsilon}$ is the site at which the wave front of the OTOC has increased past the threshold value of  $\epsilon \in (0,1)$ at time $t$, they find that $v_B$ depends on both temperature and $\epsilon$, with a slower speed of propagation for lower temperatures. We can analogously define the contour velocity
\begin{align}
    j_{c} \sim v_c^{\epsilon} \log_{10} t
\end{align}
where $j_{c}$ is the site that the contour wave front of value $\epsilon \times \tilde{s}^{\beta}$ has reached. $\tilde{s}^{\beta}$ is the equilibrium entropy density at effective temperature $\beta$ which is fixed by the energy of the quench. 

Using the cutoff  $\epsilon=0.05$, and fitting a line to the wave front in Fig.~\ref{fig:LLC2}, we arrive at a contour velocity of $v_c^{\epsilon} \sim 1.29$ (for analogous results for the disordered Heisenberg model, see Fig. \ref{fig:ed_mbl_contour} in Appendix \ref{heis_app}, where $v_c \sim 2$).
Like the butterfly velocity, this result is dependent on the choice of cutoff $\epsilon$, and may also depend on the choice of initial state. 
We are able to conclude, however, that entanglement spreading after a global quench in MBL as measured by the entanglement contour, occurs more quickly than the spreading of operators as measured by the OTOC in the disordered Heisenberg model in Ref.~\cite{2017AnP...52900318H}, where $v_B \sim 1$. More sophisticated numerical study of the entanglement contour after a global quench in MBL is warranted and could yield a more precise understanding of the relationship between $v_B$ and $v_c$ in MBL. We leave this for future work.

\begin{figure}
  \centering
  \includegraphics[width = .4\textwidth]{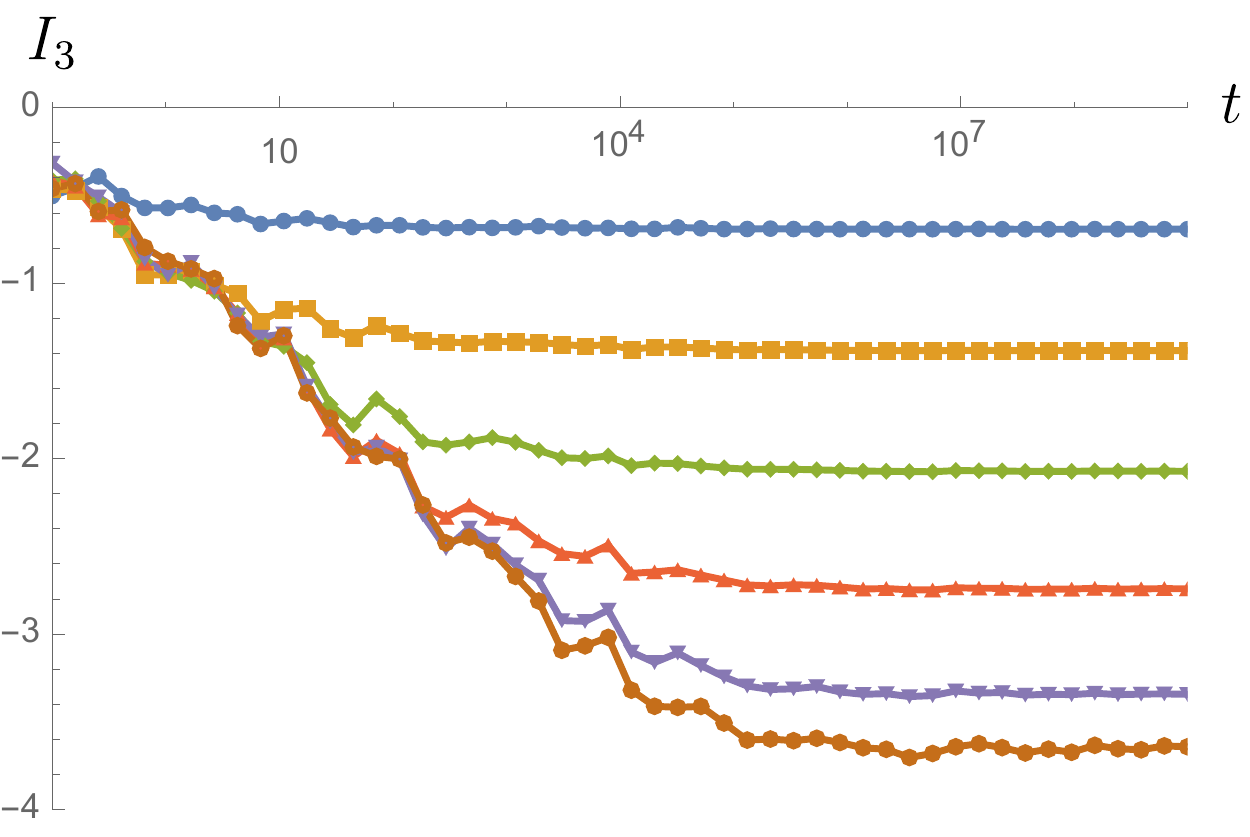}
  \includegraphics[width = .4\textwidth]{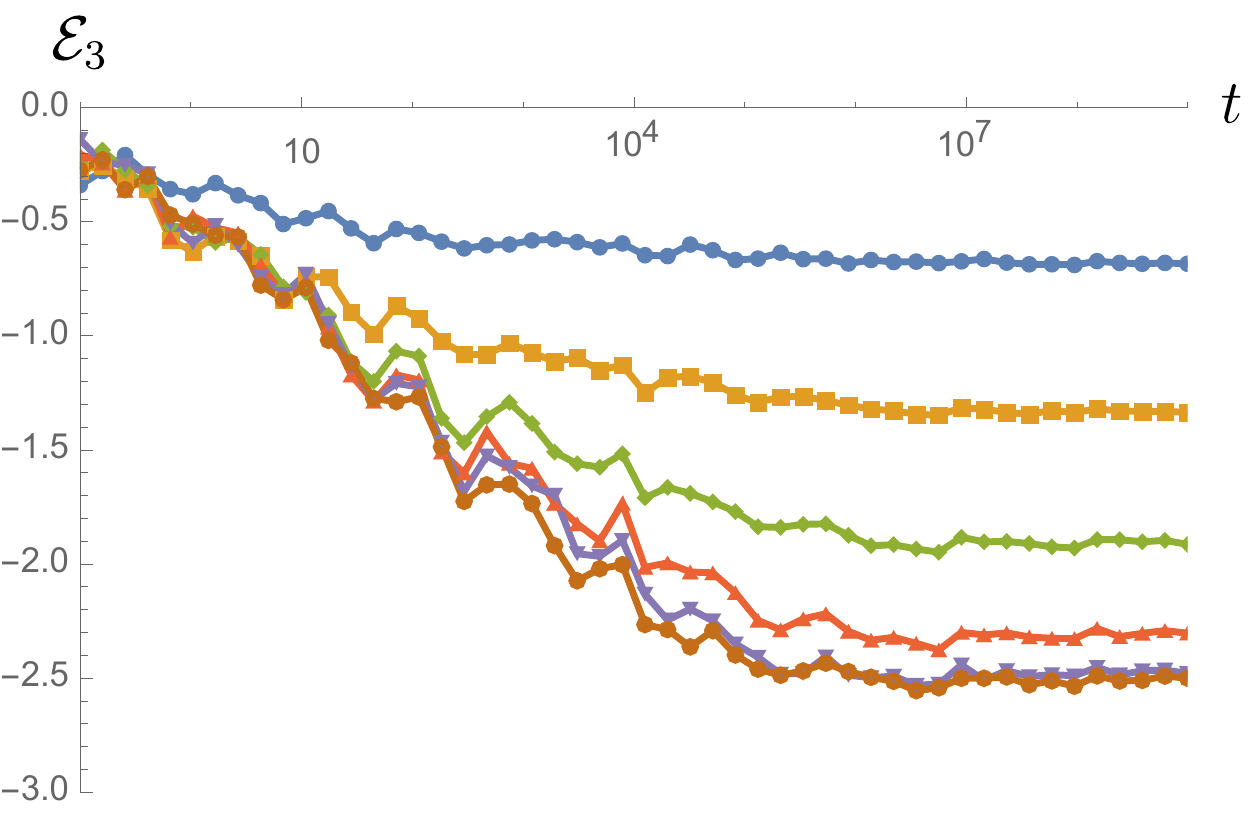}
  \caption{TOMI (top) and TOLN (bottom) for the $H_{LIOM}$ for a system of 12 input and 12 output qubits for various subinterval sizes. The disorder strength and length scale were chosen so that the LIOM numerics matched the MBL Heisenberg numerics (Fig. \ref{fig:tomi_ed}) for the 6 qubit chain. Note the logarithmic timescale and the slow saturation of both quantities, and the larger (negative) magnitude of TOMI compared to TOLN, indicating a reduced spread of quantum vs. classical information. }
  \label{fig:mbl_operator}
\end{figure}

\subsection{Multipartite Operator Entanglement}

Now that we have found that the entanglement spreads according to an emergent logarithmic light cone, we would like to understand what sort of entanglement this really is. Thermalization would necessitate this entanglement to be multipartite i.e.~the information is delocalized throughout the light cone. 

\begin{figure}
    \centering
    \includegraphics[width=.4\textwidth]{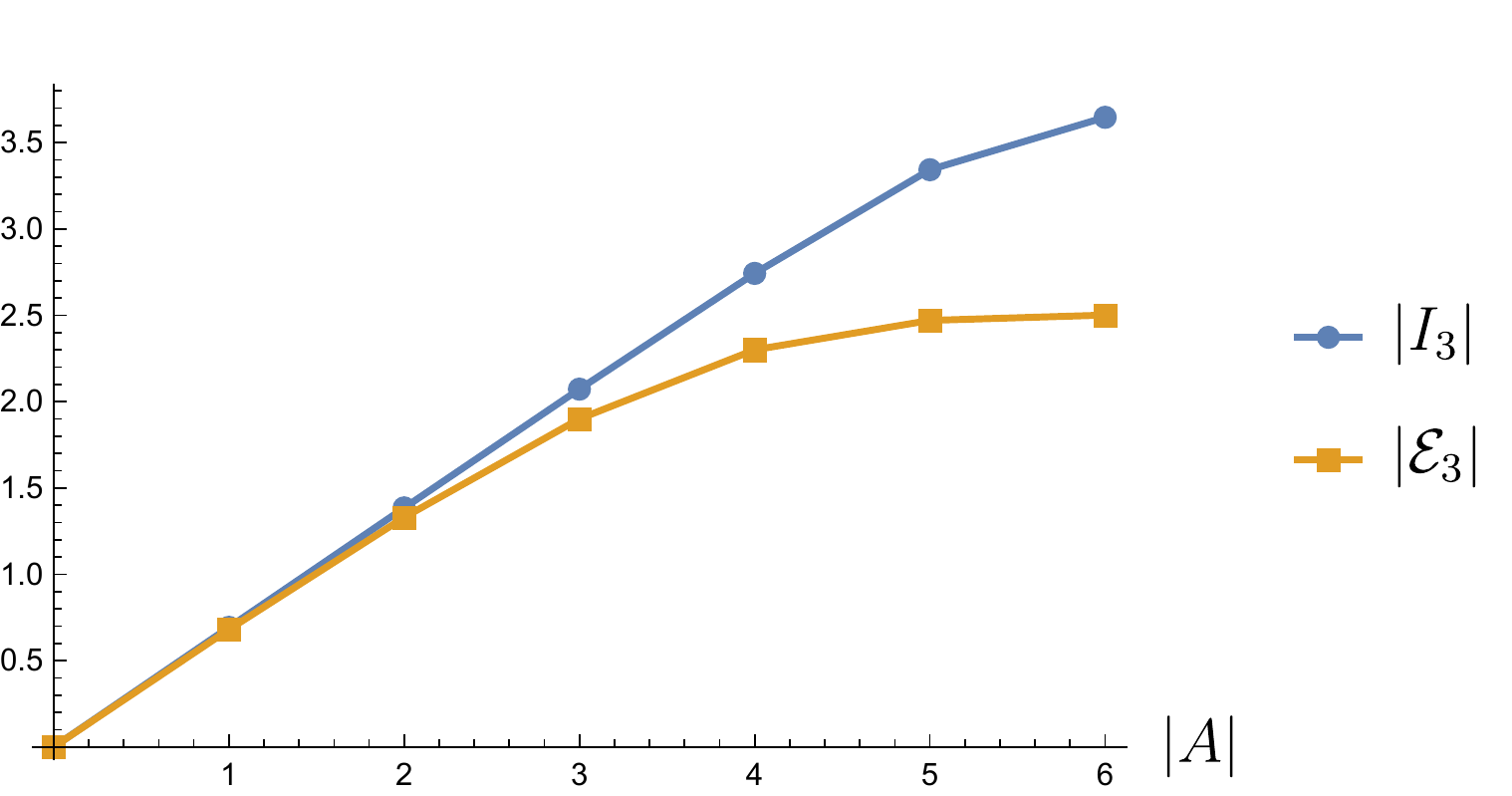}
    \caption{Late time saturation values of TOMI and TOLN for the 12 qubit MBL chain, calculated using LIOMs. The saturation values of TOMI fit well to the volume law $I_3(L_A) = \log(2) L_A$, while the TOLN values clearly follows sub-volume law scaling, indicating suppressed delocalization of quantum information.}
    \label{fig:sat_vals}
\end{figure}

We find slow scrambling in the tripartite operator mutual information and tripartite operator logarithmic negativity as seen in Fig.~\ref{fig:mbl_operator}. Like many other observables in MBL, TOMI and TOLN take an exponentially long time for the quantities to saturate. While a significant portion of the information in the input channel is delocalized under time evolution, the Haar random values of TOMI and TOLN are never reached\footnote{See Ref.~\cite{2019arXiv191014575K} for discussion on more quantum systems that scramble non-maximally. We also note that bipartite operator entanglement measures have previously been been studied in MBL systems \cite{2008PhRvB..77f4426Z,2017PhRvB..95i4206Z}.}. 
Intriguingly, it appears that the scrambled \textit{quantum information} (TOLN) may scale differently with system size than the \textit{total information} (TOMI). In Fig. \ref{fig:sat_vals}, we show the saturated values of TOMI and TOLN as a function of input interval size for our 12 site chain. While TOMI scales at or near a volume law (up until half the system size), TOLN is clearly a sub-volume law, indicating that the spreading of quantum correlations is suppressed compared to total correlations. By ``total", we are referring to both the quantum and classical (thermal) correlations to which mutual information is sensitive\footnote{See Ref.~\cite{2019arXiv190802761G} for an interesting comparison between mutual information and negativity in MBL eigenstates.} (see e.g.~Ref.~\cite{2005PhRvA..72c2317G}). Appendix \ref{heis_app} contains smaller scale operator entanglement results for the disordered Heisenberg model. To our knowledge, this is a new phenomenon that might be useful in characterizing the quasi-thermal, late-time state. It would be very interesting to further distinguish this late-time state from conventionally scrambled states. Given the small scales of our numerics, these signatures of novel scrambling behavior may be experimentally accessible in Noisy Intermediate-Scale Quantum (NISQ) devices \cite{2018arXiv180100862P} where protocols for preparation of the thermofield double state \eqref{tfd} are being developed \cite{2019PhRvL.123v0502W}. 

%
%{\color{red}
%  (https://arxiv.org/pdf/1809.04689.pdf
%  studied the negativity in MBL phase. Should we cite it?
%  Do they also contrast between quantum and classical correlations?
%  )}
%  \textcolor{blue}{[It looks like they do not contrast quantum and classical correlations. They use negativity as an extension of concurrence to higher (local) dimensional systems because it is computable. I think 1908.02761 may be more relevant to cite. I have added this citation in footnote [41].]}
%

\section{Random Singlet Phase}

In this section, we study the dynamics of a disordered free fermion model (described in Ref.~\cite{2014PhRvL.113d6802M}) that exhibits a transition between a topological and a trivial Anderson localized phase. At the critical point, the model exhibits a random singlet phase \cite{PhysRevB.50.3799} as its ground state. The RSP critical point has a number of interesting features, including CFT-like logarithmic scaling of entanglement entropy \cite{2004PhRvL..93z0602R, 2005PhRvB..72n0408L,2009JPhA...42X4010R,2011PhRvB..83d5110F}, with an effective central charge equal to $\log2$ times the central charge of the clean theory. The RSP is the fixed point of the strong disorder real space renormalization group (SDRG) \cite{PhysRevB.22.1305}, and can be seen in e.g.~the antiferromagnetic random bond Heisenberg model \cite{PhysRevB.50.3799}. It should be noted, however, that the universal features of the RSP ground state seen at the SDRG fixed point do not extend to excited states in interacting models. Indeed, while the RSP-like critical behavior extends to the excited states of a noninteracting model like the one we use here (resulting in a so-called ``quantum critical glass"\cite{PhysRevLett.114.217201, PhysRevB.93.134207}), small interactions can drive these excited states to an MBL spin glass phase. Studying the dynamics of an interacting model with a RSP ground state using the entanglement measures in this paper presents an interesting future problem. Some work in this direction has recently been done \cite{2020arXiv200104996D}, and it is found that the resulting particle-hole symmetric MBL phase exhibits entanglement growth behavior whose functional form depends on interaction strength, unlike conventional MBL.

Additional work has been done to investigate the dynamics of the random singlet phase.
For example, Ref.~\cite{2012PhRvB..85i4417I} studied the late-time growth of entanglement entropy in the RSP after a global quench using numerical methods and found it to be doubly logarithmic in time. Other works have studied entanglement growth in disordered critical phases e.g.~Refs.~\cite{PhysRevB.63.134424,2016PhRvB..93t5146Z}. We build upon this work by characterizing the spread and delocalization of information in the RSP.

We use a free fermion model with a topological phase transition that can be shown to be equivalent to the RSP via a Jordan-Wigner transformation. The model and its phase diagram are outlined in Ref.~\cite{2014PhRvL.113d6802M}. The Hamiltonian is
\begin{equation}
    H= \sum_{i} \Big \lbrack \frac{t_i}{2} \left(c_i^\dagger \left(\sigma_1 +i \sigma_2 \right)  c_{i+1} + h.c. \right) + m_i c^\dagger_i \sigma_2 c_i \Big \rbrack,
    \label{rsp_ham}
\end{equation}
where $c_i$ and $c^\dagger_i$ are two component fermions and the hopping and onsite potentials are
\begin{equation}
    t_i= 1 + \omega_i \Delta J, \quad m_i = m + \omega'_i \Delta m.
\end{equation}
Here, $\omega_i,\omega'_i \in [-0.5,0.5]$ are two independent random variables that simulate disorder, while $\Delta J$ and $\Delta m$ control the strength of the disorder. For the clean case, we have $\Delta J= \Delta m=0$. We can tune $m$ to get the topological SSH ground state for $m<1$ and a trivial gapped ground state for $m>1$. At exactly $m=1$, we have a $c=1$ critical point. When we add disorder by using nonzero values of $\Delta J$ and $\Delta m$, we change the location of the phase transition as a function of $m$. The new critical point is the random singlet phase, with effective central charge $c= \log 2$. 
% For example, setting $\Delta J= 1$ and $\Delta m=2$, the critical point appears to occur around $m=1.1$. 
In the disordered system, the topological phase becomes a localized topological phase, and the trivial gapped state becomes an Anderson insulator. 

\begin{figure}
  % [h]
  \includegraphics[width=8cm]{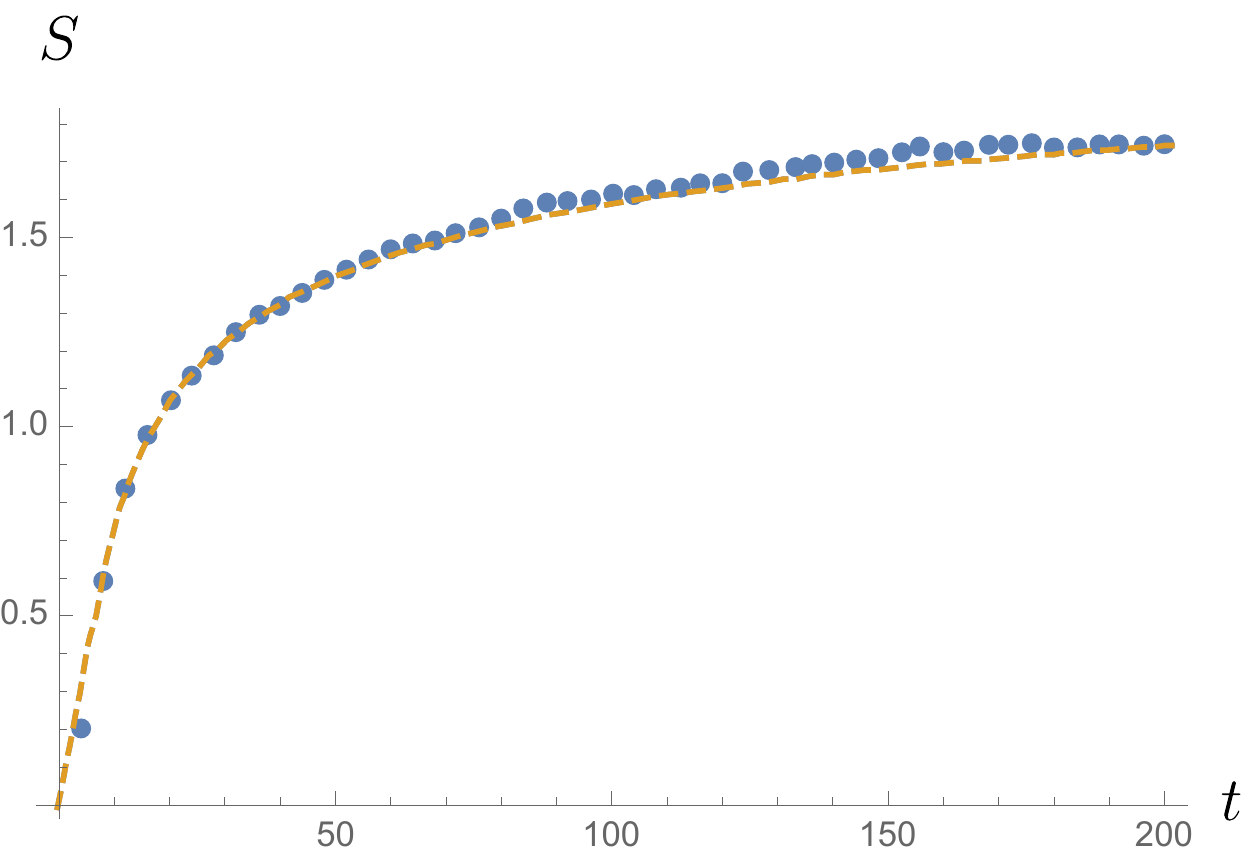}
  \caption{The half chain entanglement entropy after a global quench into the random singlet phase for a system of 100 sites. After an initial linear increase, the entropy grows as a power law 
    before transitioning to a sub-logarithmic regime. The numerical results (dots) are averaged over $3000$ disorder realizations. The analytic estimate for the entanglement entropy is displayed as a dashed line.
    We fit using $\beta_{{\it eff}}=1.06$.
    The analytic estimates derived from the quasi-particle picture show excellent agreement with the numerics.
  }
  \label{fig:RSPEEgrowth}
\end{figure}

\subsection{A Power-Law Entanglement Light Cone}

% \subsubsection{Entanglement Growth} 
\label{RSP_Entanglement}

The quench into the random singlet phase provides another example of intermediate dynamics between full localization and full thermalization. It is important that we start from a short-ranged entangled state that is not close to any eigenstates of our Hamiltonian so that we may make comparisons with the MBL results of the previous section. It is expected (and numerically verified) that other details regarding the initial state do not play an important role in the entanglement dynamics.
The initial state we choose is the half-filled ground state of (\ref{rsp_ham}) with parameters:
$
%t = 1.0,~
m = 1.8, ~
\Delta J = 0,~
\Delta m = 0.
$
This ensures that we are well inside of the gapped, disorderless phase where correlation lengths are short, so that entanglement is short-ranged. At time $t=0$, we quench to the random singlet phase by instantaneously changing the parameters to
$
%t = 1.0,~
m = 1.1, ~\Delta J = 1,~\Delta m = 2.
$

The entanglement entropy growth is depicted in Fig.~\ref{fig:RSPEEgrowth}. Qualitatively, we see an initial linear growth, followed by what appears to be a power law growth, which eventually settles to a very slow, sub-logarithmic growth. In order to obtain an analytical estimate for the entanglement dynamics, we use the quasi-particle picture that is applicable to integrable systems \cite{2005JSMTE..04..010C,2006PhRvL..96m6801C,2017PNAS..114.7947A,2018ScPP....4...17A}.
The master formula is
\begin{equation}
  S(t) \propto t \int_{|v(\varepsilon)| t < \ell} d\varepsilon v(\varepsilon) f(\varepsilon)
  + \ell \int_{|v(\varepsilon)| t > \ell}d\varepsilon f(\varepsilon),
    \label{qparticle_ent1}
\end{equation}
where $\ell$ is the length of the interval, $v(\varepsilon)$ is the velocity of the quasi-particles at energy $\varepsilon$ and $f(\varepsilon)$ is the entanglement production rate of quasi-particles. In other words, $f(\varepsilon)$ is the extent to which each mode contributes to the entanglement entropy. For this function, we can use the entropy of each occupied fermionic mode
\begin{equation}
    f(\varepsilon)= -(1-n(\varepsilon)) \log ( 1- n(\varepsilon)) - n(\varepsilon) \log n(\varepsilon),
\end{equation}
where $n(\varepsilon)$ is the occupation number of each mode after the quench. We use the Fermi-Dirac distribution 
\begin{equation}
    n(\varepsilon) = \frac{1}{1+e^{\beta_{ {\it eff}} \varepsilon}},
\end{equation}
which provides an excellent approximation.
Here, $\beta_{{\it eff}}$ is an effective inverse temperature, determined by the energy of the initial state.

Using the density of states for the SDRG fixed point,
$\rho(\varepsilon)$ 
\cite{PhysRev.92.1331},
we can compute the velocity of the associated quasi-particles 
\begin{equation}
    \rho (\varepsilon) = \frac{\rho_0}{\varepsilon |\log \varepsilon |^3} \quad \rightarrow \quad v(\varepsilon) =\frac{\varepsilon |\log \varepsilon |^3}{\rho_0}.
    \label{qparticle_ent2}
\end{equation}
The above density of states is quite unusual, though we have verified it numerically, reassuring us that we are closely approximating the infinite disorder fixed point. It displays a concentration of low energy ``slow" modes between $\varepsilon=0$ and $\varepsilon=1$. These may be responsible for the long-time growth of entanglement entropy. It should be emphasized, however, that the above density of states comes from the fixed point of a real space RG procedure, and is only expected to be valid asymptotically as $\varepsilon \rightarrow 0$. Using the standard form for the semiclassical particle velocity, 
$v(\varepsilon)={d \varepsilon(k)}/{dk}\big |_{k(\varepsilon)}$, which we have done, is also not exactly correct, since the eigenstates of the disordered model are not labeled by momentum $k$. Disorder averaging, however, restores approximate translational symmetry, and the above form of the quasi-particle velocities yields results consistent with numerics. The remaining details of this calculation can be found in Appendix \ref{RSP_ent_calc}. The analytical form of the entanglement entropy is very complicated and not terribly enlightening, but agrees remarkably well with the numerical results as shown in Fig.~\ref{fig:RSPEEgrowth}.

\begin{figure}
    \centering
    \includegraphics[width=8cm]{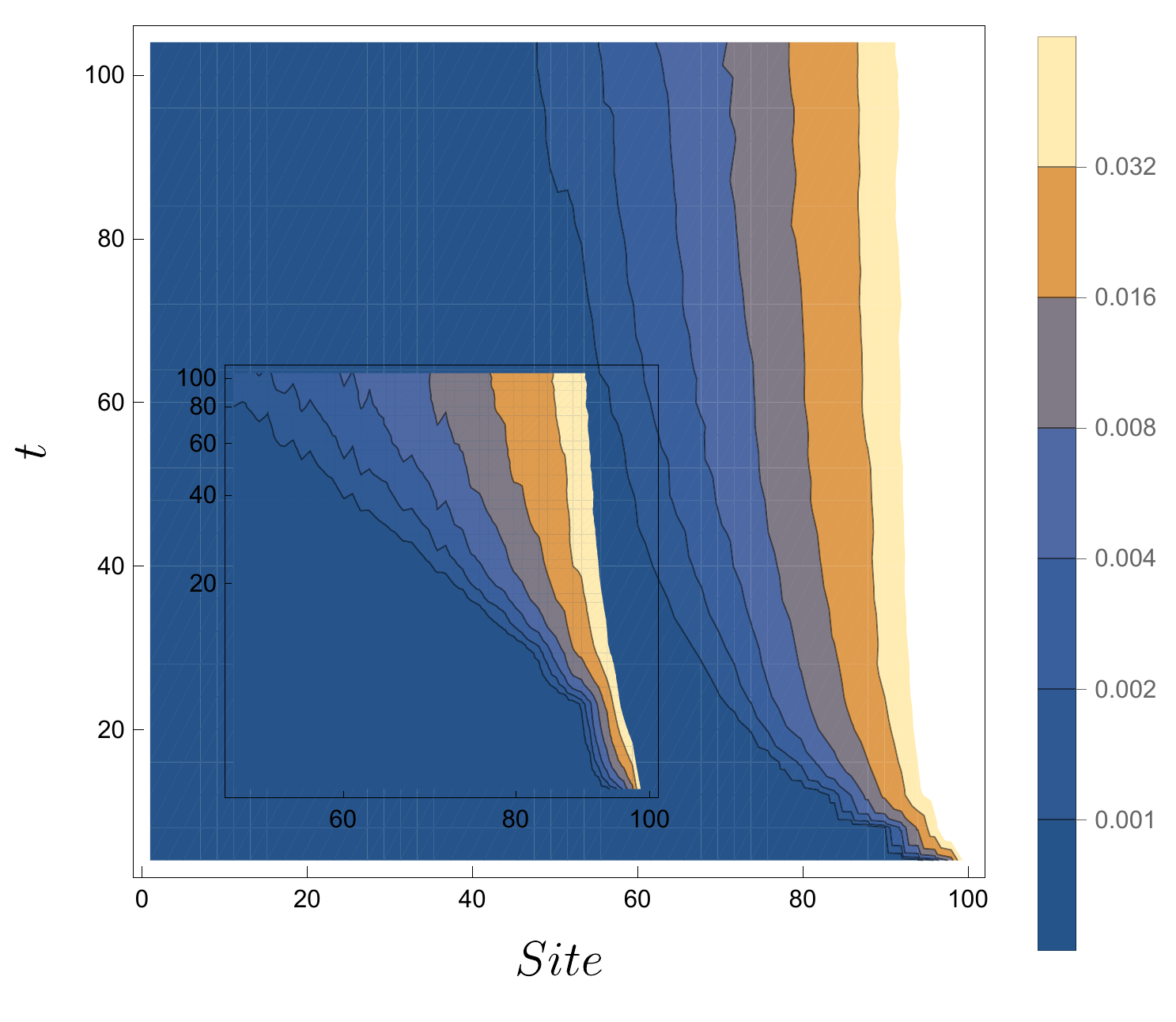}
    \caption{
    The entanglement contour for the leftmost $100$ sites in an open chain of
    $200$ stes averaged over $500$ disorder realizations and normalized by $\log
    2$. We use $\Delta J= 1.0,~ \Delta m= 2.0$, and $m=1.1$.
    We see the emergence of a power-law light cone for the entanglement contour, approximately following $x_c \sim t^\alpha$, with $\alpha \sim 0.264 $. If we tune $m$ below or above this critical point, we observe a completely localized entanglement contour profile in the topological and Anderson phases, respectively. 
    The inset shows the same data on a $\log$-$\log$ scale to make the power-law light cone manifest. 
    }
    \label{fig:rsp_quench}
\end{figure}

We now use the contour to investigate finer grained aspects of the entanglement entropy growth. 
The result of a global quench from a gapped ground state at half filling can be
seen in Fig.~\ref{fig:rsp_quench}. Though the production of entanglement is
weaker than it is in the clean limit, the contour demonstrates nontrivial
spreading, and carves out a novel (sub-linear) power-law light cone.
From the numerical fit,
we determine the power as 
\begin{align}
x_c \sim t^\alpha \quad \mbox{with}\quad \alpha \sim 0.264. 
\end{align}
This power-law behavior sits between the ballistic spreading of entanglement in clean systems and the logarithmic spreading in MBL. It is important to note that we are only probing relatively short times in these numerics, and that the power-law lightcone is likely a transient effect. Still, it is interesting to see how the entanglement contour captures this early time behavior.

\subsection{Multipartite Operator Entanglement}

While the RSP has nontrivial entanglement spreading as demonstrated in the previous section, we still need to determine what sort of entanglement is generated. In particular, is this entanglement multipartite, leading to thermalization of subsystems at late times? Our expectation is that bipartite entanglement should dominate, since our model is free. Indeed, this is what we find when computing the operator entanglement. We include the details of the free fermion operator entanglement computation in Appendix \ref{op_app}.
The BOMI and BOLN are consistent with the quasi-particle picture with a non-trivial dispersion relation.
The tripartite mutual information is negative and appears to saturate at some value that is independent of the subsystem size. Naively, we would then be led to believe that there is some amount of delocalization of information. However, we determine that this nontrivial $I_3$ is merely a finite size effect, which disappears as we go deeper into the RSP by increasing disorder strength. 

\section{Discussion}

In summary, we have investigated the non-equilibrium dynamics of two distinct non-thermalizing phases; a local integral of motion model of many-body localization, and (a free fermion realization of) the random singlet phase. 

\paragraph*{Contour velocity vs.~butterfly velocity}
Calculating the entanglement contour after a global quench revealed a logarithmic light cone of entanglement spreading in MBL. This light cone was similar, but not identical to the logarithmic light cone seen for the OTOC. Meanwhile, in the RSP, the entanglement contour yielded a novel power-law light cone, despite trivial spreading of the OTOC in that system (not shown). Each light cone defines a velocity and we now comment on how these may be related or distinguished. Following a global quantum quench, the entanglement contour propagates from the entangling surfaces. The contour velocity is the speed at which the wave front propagates. One must impose some cutoff value of the contour in order to define the wave front. On the other hand, the butterfly velocity corresponds to the speed at which local operators spread, in contrast to the contour velocity which corresponds to the speed at which correlations spread. The butterfly velocity is defined through the out-of-time-ordered correlator (OTOC). Naturally, the contour velocity and butterfly velocity will be related. They are, in general, different speeds partially because the OTOC only time evolves one of the operators in the correlation function, while the entanglement contour probes time evolved \textit{states} i.e.~all operators are time evolved. Because all operators have been given the chance to spread in time, the contour velocity will be roughly twice as fast as the butterfly velocity. In 2d conformal field theories, it is true that $v_c = 2 v_B$, and it appears to be approximately true (in logarithmic time) for the MBL Heisenberg model. We believe it would be interesting to study the relationship between $v_c$ and $v_B$ in more general systems.

\paragraph*{Late-time quasi-equilibration}
To characterize the kind of entanglement generated by time evolution, we calculated operator mutual information and negativity for both phases. We found that the many-body localized system demonstrated slow, but nontrivial saturation of tripartite operator mutual information and negativity, to values smaller in magnitude than the Haar-random case. The late-time values of TOLN were even further suppressed. This indicates a level of weakly-scrambled quasi-equilibration in MBL. However, the late-time state is clearly not fully thermalized because it still retains memory of the initial LIOMs. We wonder whether this is a different sort of ``thermalized" quasi-equilibrium state and whether it be characterized by a generalized Gibbs ensemble or a generalization of the so-called canonical thermal pure quantum (cTPQ) states \cite{2018NatCo...9.1635N}. 

In contrast to the MBL phase, we found the RSP to demonstrate multipartite operator measures that decayed to zero with increasing disorder, indicating that despite nontrivial entanglement spreading, the RSP does not delocalize quantum information. This is unsurprising, as we used a free fermion model for the RSP. Our results indicate a broad range of behavior of state and operator dynamics lying between clean, free particle systems and the maximally scrambling holographic or Haar-random systems. Other intermediate systems that may be worth investigating include models with quasiperiodic potentials (in particular the Aubry-Andr\'e model \cite{Aubry1980AnalyticityBA}), Floquet systems, and random unitary circuits with measurements\footnote{See e.g.~Refs.~\cite{2019ScPP....6...50M, 2019arXiv190207199X, 2014PhRvL.112o0401L,2019arXiv190805289F,2018PhRvB..98t5136L, 2019PhRvX...9c1009S, 2019arXiv190808051J} for examples of recent work in this direction.
}.

As entanglement measures become more experimentally accessible, the ubiquity of disorder in physical systems could make slowly scrambling systems an interesting testbed for quantum information dynamics in the lab. Many-body localization has been realized experimentally in several different settings, including superconducting qubits \cite{2017arXiv170907108R} and optical lattice systems \cite{2015Sci...349..842S}. Information theoretic measures, for example the quantum Fisher information \cite{2016NatPh..12..907S} and the second R\'enyi entanglement entropy \cite{ 2019Sci...364..256L, 2019Sci...364..260B}, have become measurable in the lab. As experiments continue to improve, we hope to see the finer-grained probes of entanglement spreading and scrambling that we have studied measured experimentally in MBL and other disordered systems.

%%%%%%%%%%%%%%%%%%%%%%%%%%%%%%%%%%%%%%%%%%%%%%%%%%%%%%%

\acknowledgements{We thank Dmitry Abanin, Masahiro Nozaki, Hassan Shapourian, and Masaki Tezuka for useful comments. SR is supported by a Simons Investigator Grant from the Simons Foundation.}
This work was supported in part by the National Science Foundation grant DMR
2001181.

\paragraph*{Note added}
During the completion of this work, a paper appeared that also studies operator entanglement in MBL systems using an effective Hamiltonian \cite{2020arXiv201214609M}. This Hamiltonian is distinct from the one we study in Section \ref{MBL_sec}.

\appendix

\section{Disordered Heisenberg MBL Results }
\label{heis_app}

Here we present supplementary entanglement contour and operator entanglement results for the many-body localized disordered Heisenberg model, which has the following Hamiltonian.

\begin{equation}
    H= J \sum_i \left( \sigma^x_i \sigma^x_{i+1} + \sigma^y_i \sigma^y_{i+1} + \sigma^z_i \sigma^z_{i+1}\right) + \sum_i h_i \sigma^z_i,
    \label{HeisHam}
\end{equation}
where $h_i$ is a random variable from the uniform distribution $[-h,h]$.
This model is believed to be fully many-body localized for $J=1$ and $h\gtrsim7$. As in Ref.~\cite{2017AnP...52900318H}, we will use $h=16$ to ensure a short localization length.

\begin{figure}
    \centering
    \includegraphics[width=.4\textwidth]{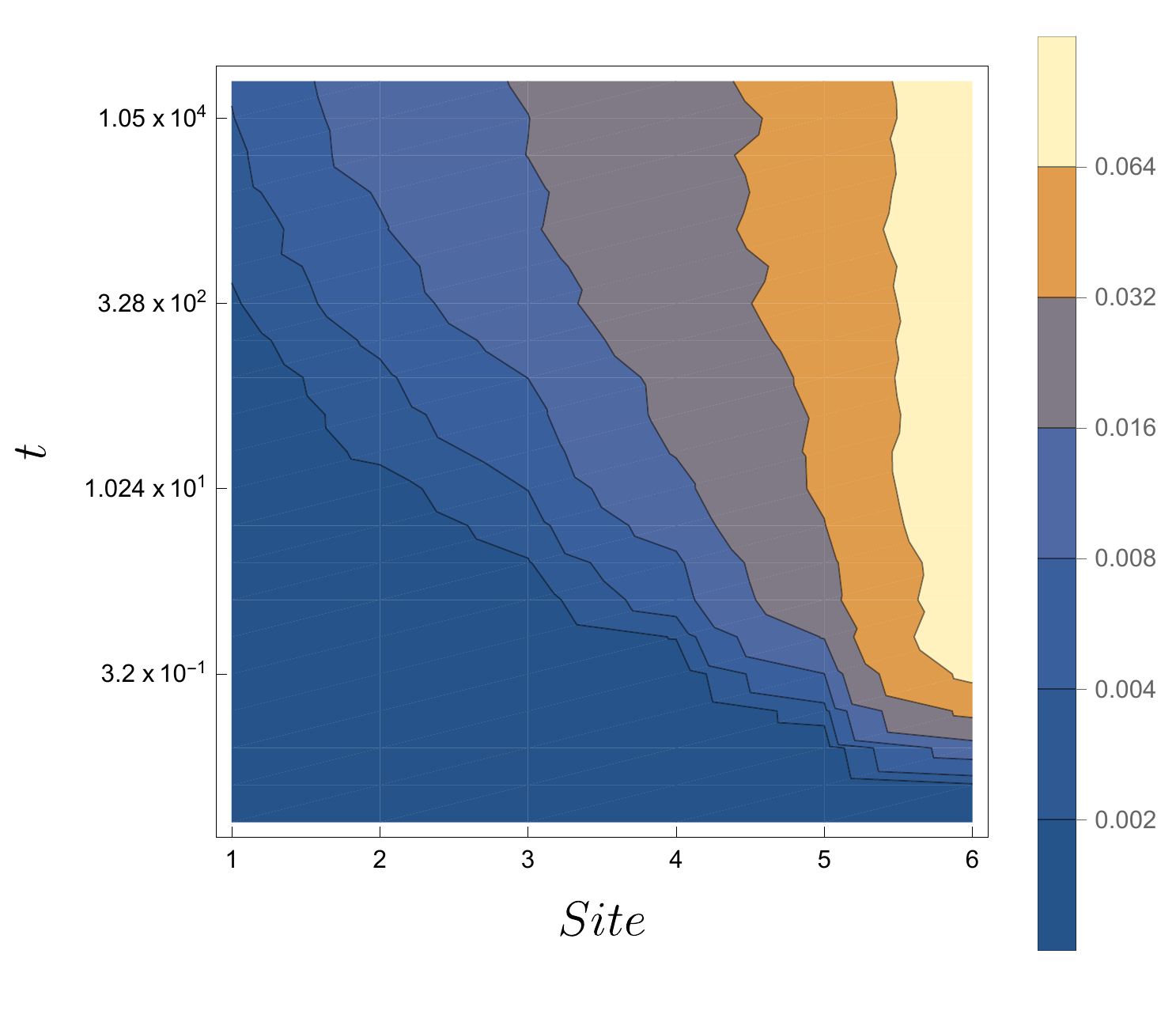}
    \caption{The entanglement contour for the leftmost six site subinterval (divided by $\log2$) in the MBL Heisenberg model after quenching from random product states. The results were averaged over 400 disorder realizations.}
    \label{fig:ed_mbl_contour}
\end{figure}

We first present the entanglement contour, in Fig. \ref{fig:ed_mbl_contour}, which, though demonstrating slightly different features than the LIOM lightcone in Fig. \ref{fig:LLC2}, retains the essential logarithmic lightcone, this time with a contour velocity of $v_c \sim 2$ if measured along the $0.002$ contour. Here, the contour velocity appears to depend on the choice of cutoff, and the overall magnitude of entanglement is lower. This may be a model-dependent effect.

\begin{figure}
    \centering
    \includegraphics[width=.4\textwidth]{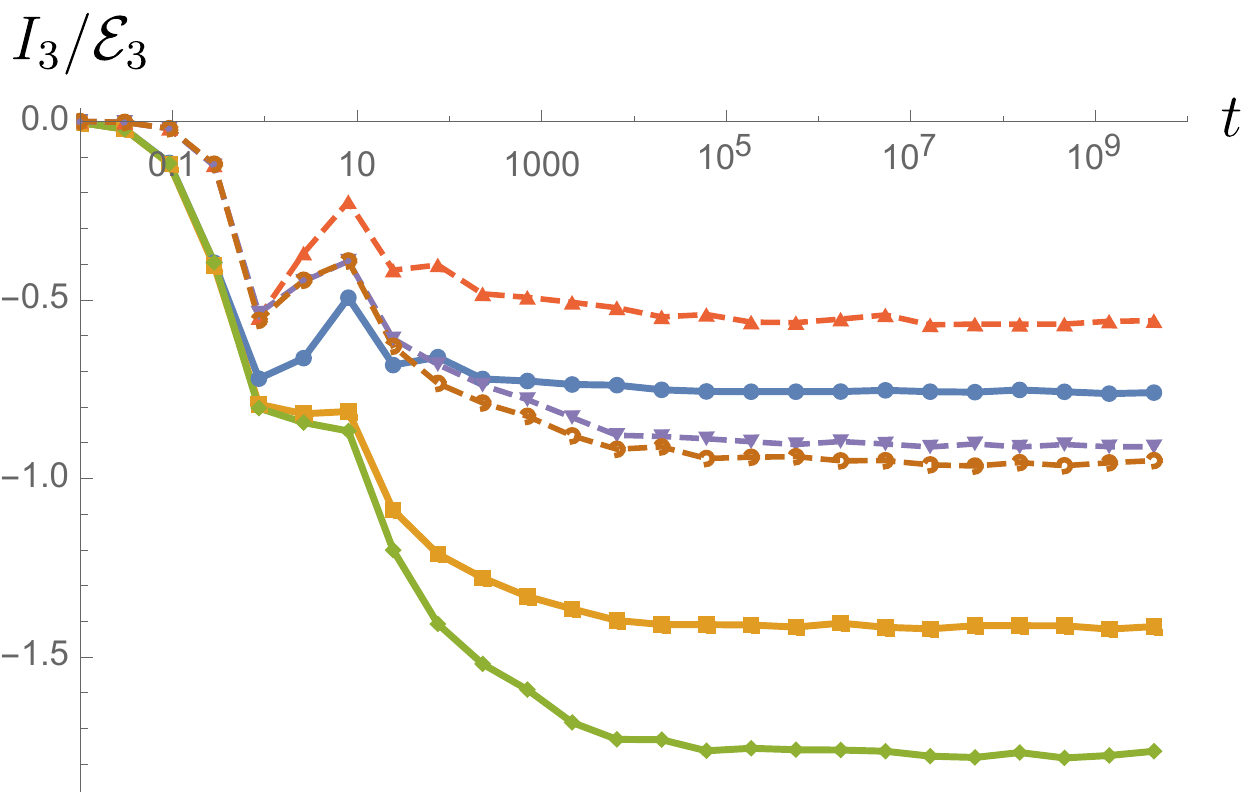}
    \caption{TOMI (solid lines) and TOLN (dashed lines) computed for a six site random Heisenberg model for input intervals of one, two and three sites. As in the case of LIOM calculation, both values saturate exponentially slowly, and the larger (negative) magnitude of the TOMI saturation value suggests a suppression of quantum information spreading.}
    \label{fig:tomi_ed}
\end{figure}

Next, we present the exact diagonalization results of the operator mutual information and negativity in the disordered Heisenberg model. Because we must directly compute and partial trace over the operator-state density matrix, we are limited to a system size of six spins. The results for TOMI and TOLN are depicted in Fig. \ref{fig:tomi_ed}. These results were computed before the LIOM results, and the dramatic difference in the scaling of TOMI and TOLN motivated further calculations using LIOMs.

\section{Computing Operator Entanglement and Negativity with MBL Integrals of Motion}
\label{LIOM_calc}

We will use the LIOM basis to compute operator mutual information for a MBL system. We start with the following Hamiltonian,
\begin{equation}
    H_{{\it LIOM}} = - \sum_i h_i \sigma^z_i - \sum_{i<j} J^{(2)}_{ij}  \sigma^z_i \sigma^z_j- \sum_{i<j<k} J^{(3)}_{ijk}  \sigma^z_i \sigma^z_j\sigma^z_k +
    \cdots
\end{equation}
where we have truncated terms beyond third order. In the LIOM basis, the time evolution operator for a chain of $N$ spins is
\begin{align}
&
U = \sum_{ \{s\} } \exp \left[ 
i t 
H_{{\it LIOM}}(s)
%    \left(  \sum_i h_i s_i  +\sum_{i<j} J^{(2)}_{ij}  s_i s_j +\sum_{i<j<k} J^{(3)}_{ijk}  s_i s_j s_k \right)  
\right] 
|s_1 \cdots s_N \rangle \langle s_1 \cdots s_N|,
\nonumber \\
&
H_{{\it LIOM}}(s)
=
\sum_i h_i s_i  +\sum_{i<j} J^{(2)}_{ij}  s_i s_j +\sum_{i<j<k} J^{(3)}_{ijk}  s_i s_j s_k,
\end{align}
where $\sum_{\{ s \}}$ is a sum over all $2^N$ classical spin configurations. Since $U$ is already diagonal in the LIOM basis, and we can replace the Pauli operators in the Hamiltonian with classical spin values. To compute operator entanglement measures we use the channel-state duality to obtain the $U$ operator state, and then we take the outer product to get the density matrix of the doubled Hilbert space state
\begin{gather}
\rho_U = \sum_{ \{s\},\{s'\} } \exp \left[ i t \left( 
H_{{\it LIOM}}(s) - H_{{\it LIOM}}(s')\right)  \right] \\
    \times |s_1 \cdots s_N \rangle \langle s'_1 \cdots s'_N|  \nonumber.
\end{gather}
We can go further by subdividing the input and output Hilbert spaces into intervals $A$, $D$ and $B$, $C$, respectively:
\begin{gather}
    \rho_U = \sum_{ \{s\}, \{s'\} } \exp \left[ i t \left( H_{LIOM}(s) - H_{LIOM}(s')\right)  \right] \\
    \times | \{s\}_A, \{s\}_B, \{s\}_C,\{s\}_D  \rangle \langle \{s'\}_A, \{s'\}_B, \{s'\}_C,\{s'\}_D |\nonumber.
\end{gather}

We choose $A$ and $B$ to be of the same size and position in their respective Hilbert spaces. To obtain a reduced density matrix for the operator state --- for example for $A\cup B$ ---  we can trace out $C \cup D$
\begin{gather}
    \rho_{AB} = \sum_{ \{s\}_A, \{s\}_B,\{s'\}_A, \{s'\}_B } \sum_{ \{s\}_C, \{s\}_D} \\
    \times \exp \left[ i t \left( H_{LIOM}(s_A,s_B,s_C,s_D) - H_{LIOM}(s'_A,s'_B,s_C,s_D)\right)  \right]  \nonumber \\
    \times | \{s\}_A, \{s\}_B \rangle \langle \{s'\}_A, \{s'\}_B |\nonumber.
\end{gather}
By performing the sums over $s_C$ and $s_D$ to calculate the matrix elements of $\rho_{AB}$, we can avoid directly storing and tracing over a $2^{2N} \times 2^{2N}$ matrix, and instead deal with, at most, a $2^{N} \times 2^{N}$ matrix. This is not a dramatic decrease in numerical overhead, but it allows us to (nearly) double the size of the system in question when computing operator entanglement measures, as compared to directly performing the partial trace over the full operator state density matrix.

\section{Operator Entanglement and Negativity for Free Fermions} \label{op_app}
 In this appendix, we present a review of how to compute operator entanglement
 measures for free fermion systems using the correlator method.

\subsection{Operator State for Free Fermions}
We wish to compute the operator mutual information for the following state
\begin{equation}
  \ket{U_\beta(t)}=e^{-\frac{i t}{2}(H_1+H_2)}
  \ket{{\it TFD}_\beta}
\end{equation}
for a free fermion Hamiltonian with no superconducting terms
\begin{equation}\label{XXchain}
\hat{H} = \sum_{i,j=1}^L H_{i,j} c_i^\dagger c_{j}.
\end{equation}
We can diagonalize this Hamiltonian with
a unitary matrix $U$, so $H=U D U^\dagger$:
\begin{equation}
  \hat{H}= \underbrace{c_i^\dagger U_{ik}}_{\equiv \psi_k^\dagger}
  D_{kl}\underbrace{U_{lj}^\dagger c_j}_{\equiv \psi_l}
  =\psi_k^\dagger D_{kl} \psi^{\ }_l=\epsilon_k \psi_k^\dagger \psi^{\ }_k.
\end{equation}
We then write down the thermofield double state with the fermions in the diagonal basis.
\begin{align}
  \ket{{\it TFD}_\beta}
  &= \frac{1}{\sqrt{Z}}\prod_k
    \Big(
    \sum_{i_k}e^{-\frac{\beta}{2}\epsilon_k \psi_k^\dagger \psi^{\ }_k}
    \ket{i_k}\ket{i_k^*} \Big)
  \nonumber \\
% &= \frac{1}{\sqrt{Z}}\prod_k \left(\ket{0}_k+e^{-\frac{\beta}{2}\epsilon_k}\hat{\psi}_{Ak}^\dagger\hat{\psi}_{Bk}^\dagger\ket{0}_k \right)\\ \nonumber
&=\frac{1}{\sqrt{Z}}\prod_k \left(1+e^{-\frac{\beta}{2}\epsilon_k}{\psi}_{Ak}^\dagger{\psi}_{Bk}^\dagger \right) \ket{0}.
\end{align}
Requiring 
$\langle {\it TFD}_\beta|{\it TFD}_\beta\rangle = 1$
fixes the normalization factor as
$
Z= \prod_k \left(1+e^{-\beta \epsilon_k} \right)
$.
The normalized thermofield double state is thus
\begin{align}
  \ket{{\it TFD}_{\beta}}
  &=  \prod_k
    \Bigg(
    \underbrace{\frac{e^{\frac{\beta}{2} \epsilon_k}}{\sqrt{1+e^{\beta \epsilon_k}}}}_{\equiv \cos \theta_k}+\underbrace{\frac{1}{\sqrt{1+e^{\beta \epsilon_k}}}}_{\equiv \sin \theta_k} {\psi}_{Ak}^\dagger {\psi}_{Bk}^\dagger
    \Bigg)\ket{0}
\end{align}
and the time evolved operator state becomes
\begin{align}\label{OperatorState}
  \ket{U(t)}
%  &= e^{-\frac{i t}{2}(\hat{H}_A+\hat{H}_B)} \ket{{\it TFD}_\beta}
%              \nonumber \\ 
              % \nonumber
              % &= e^{-\frac{i t}{2}(\hat{H}_A+\hat{H}_B)}\prod_k \left(\cos \theta_k+\sin\theta_k \hat{\psi}_{Ak}^\dagger \hat{\psi}_{Bk}^\dagger \right)\ket{0} \\
   &= \prod_k\left(\cos \theta_k+\sin \theta_k e^{-i t \epsilon_k}
     {\psi}_{Ak}^\dagger {\psi}_{Bk}^\dagger\right) \ket{0}.
\end{align}

\paragraph*{Alternate form of Operator State}
We now rewrite the operator state \eqref{OperatorState} by using the holes of the $B$ Hilbert space instead of the particles because this allows us to use the regular correlation matrix without pairing terms. Let $\chi_{Ak}$, $\chi_{Bk}$ be new fermion operators and consider
\begin{align}
  &
    \prod_k \left(\cos \theta_k \chi_{Bk}^\dagger + \sin \theta_k e^{-it \epsilon_k}\chi_{Ak}^\dagger \right) \ket{0}_\chi 
\nonumber \\
  &
    = \prod_k \left(\cos \theta_k
    + \sin \theta_k e^{-it \epsilon_k}\chi_{Ak}^\dagger\chi^{\ }_{Bk}
    \right)\prod_q \chi_{Bq}^\dagger \ket{0}_\chi. 
\label{alternate_state}
\end{align}
We now define $\chi_{Ak}=\psi_{Ak}$, $\chi_{Bk}=\psi_{Bk}^\dagger$
and note $\ket{0}_\psi \sim \prod_q \chi_{Bq}^\dagger \ket{0}_\chi$
because $\psi_{Bp}\ket{0}_\psi \sim\chi_{Bp}^\dagger \prod_q \chi_{Bq}^\dagger
\ket{0}_\chi = 0$, since $(\chi^\dagger)^2=0$.
The $\psi$ and $\chi$ fermions are related
by a particle-hole transformation on $H_B$.
The state \eqref{alternate_state} is already normalized. 
In terms of the $\chi_{Ik}$ fermions, we have
\begin{equation}\label{OperatorStateChi}
\ket{U(t)} = \prod_k (\cos \theta_k \chi_{Bk}^\dagger+\sin \theta_k e^{-it \epsilon_k}\chi_{Ak}^\dagger)\ket{0}
\end{equation}
where it is understood that $\ket{0} = \ket{0}_\chi$. 

\subsection{Correlator Method}
Now we compute the correlation matrix
\begin{equation}
\bra{U(t)} \chi_{Ik}^\dagger \chi^{\ }_{J k'}\ket{U(t)}
= \delta_{kk'}\bra{U(t)} \chi_{Ik}^\dagger \chi^{\ }_{J k}\ket{U(t)},
\end{equation}
where we noted that
%Note that
if $k\neq k'$,
$\chi_{Ik}^\dagger \chi^{\ }_{Jk'} = -\chi^{\ }_{Jk'}\chi_{Ik}^\dagger$
annihilates $\ket{U(t)}$.
%This leads to the simplification
%\begin{equation}
%  \bra{U(t)} \chi_{Ik}^\dagger \chi^{\ }_{J k'}\ket{U(t)}
%  = \delta_{kk'}\bra{U(t)} \chi_{Ik}^\dagger \chi^{\ }_{J k}\ket{U(t)}.
%\end{equation}
Suppose that the product over momenta in \eqref{OperatorStateChi} is arranged in
increasing order.
The matrix element for each momentum $k$
can be computed from
\begin{align}
  &
    \bra{U(t)}  \chi_{Ik}^\dagger \chi^{\ }_{J k}\ket{U(t)} 
=  \bra{0} \prod_{q>k} (\cos \theta_q \chi_{Bq}+\sin \theta_q e^{i t \epsilon_q}\chi_{Aq})  
\nonumber \\
  &\quad
    \times\Big[ (\cos \theta_k \chi_{Bk}+\sin \theta_k e^{i t \epsilon_k}\chi_{Ak}) \chi_{Ik}^\dagger \chi_{Jk}
\nonumber \\
  &
    \qquad 
    \times
    (\cos \theta_k \chi_{Bk}^\dagger
    +\sin\theta_k e^{-it \epsilon_k}\chi_{Ak}^\dagger) \Big] 
     \nonumber \\
  &
    \quad 
  \times
    \prod_{p>k}(\cos \theta_p \chi_{Bp}^\dagger+\sin\theta_p e^{-it \epsilon_p}\chi_{Ap}^\dagger)\ket{0}
\end{align}
for the four possible values of $(I,J)$.
%Let us compute the term in the square bracket
%for the four possible values of $(I,J)$.
%We drop all terms that vanish when inserted in the correlator.
%\begin{align}
%I=A, J=A:&
%\quad
%\sin^2 \theta_k,
%\nonumber \\
%I=A,J=B:&
%\quad
%\sin \theta_k\cos \theta_k e^{i t \epsilon_k},
%\nonumber \\
%I=B,J=A:&
%\quad
%\sin \theta_k\cos \theta_k e^{-i t \epsilon_k},
%\nonumber \\
%I=B,J=B:&
%\quad
% \cos^2 \theta_k.
%\end{align}
The correlation matrix in momentum space is thus
\begin{align}
\bra{U(t)}& \chi_{Ik}^\dagger \chi^{\ }_{J k'}\ket{U(t)} = 
\\
&\delta_{k k'}
\begin{pmatrix}
\sin^2 \theta_k & \sin\theta_k \cos\theta_k e^{i t \epsilon_k}  \\ \nonumber
\sin\theta_k \cos\theta_k e^{-i t \epsilon_k} & \cos^2\theta_k
\end{pmatrix}.
\end{align}
The correlation matrix in real space,
which we need for the calculation of the entanglement entropy,
is given by
\begin{align}
  & \bra{U(t)} \chi_{Ix}^\dagger \chi^{\ }_{J x'}\ket{U(t)} \\ \nonumber
  &= \sum_k V_{xk}^*
    \begin{pmatrix}
      \sin^2 \theta_k & \sin\theta_k \cos\theta_k e^{i t \epsilon_k}  \\ \nonumber
      \sin\theta_k \cos\theta_k e^{-i t \epsilon_k} & \cos^2\theta_k
    \end{pmatrix}
                                                      V^t_{kx'},
\end{align}
where a general position space Hamiltonian
can be daigonalized through the unitary matrix as,
\begin{align}
\hat{H} &= \sum_{x,y}\chi_x^\dagger H_{xy}\chi_y 
= \sum_{x,y}\underbrace{\chi^\dagger_x V_{xk}}_{\chi_k^\dagger}\underbrace{D_{kq}}_{\epsilon_k \delta_{kq}}\underbrace{V^\dagger_{qy}\chi_y}_{\chi_q} 
          \nonumber \\
&= \sum_k \epsilon_k \chi^\dagger_k \chi^{\ }_k,
\end{align}

Finally,
the von Neumann entropy is given by
\begin{equation}
S(t)= - \sum_k \left[\xi_k(t) \ln \xi_k(t)+(1-\xi_k(t))\ln (1-\xi_k(t)) \right],
\end{equation}
where $\xi_k(t)$ are the eigenvalues of the correlation matrix truncated to the entries corresponding to degrees of freedom in our subsystem.

\section{Quasi-particles for the RSP} \label{RSP_ent_calc}

In this appendix, we present the details in the derivation of the analytical estimate for the entanglement entropy of the RSP after a global quench described in the main text. Combining the ingredients from \eqref{qparticle_ent1}-\eqref{qparticle_ent2}, we obtain the integral
\begin{align}
    S(t) &\propto \frac{t}{\rho_0} \left[\int_{|v(\varepsilon)| t < \ell} d\varepsilon \varepsilon | \log \varepsilon |^3 +\ell \int_{|v(\varepsilon)| t > \ell}d\varepsilon \right]
    \nonumber
    \\
    &\qquad 
    \times \left [ \log ( 1+ e^{\beta \varepsilon}) - \frac{\beta \varepsilon e^{\beta \varepsilon}}{1+e^{\beta \varepsilon}} \right]
    % \nonumber
    % \\ 
    % &+ \left ( \log ( 1+ e^{\beta \varepsilon}) - \frac{\beta \varepsilon e^{\beta \varepsilon}}{1+e^{\beta \varepsilon}} \right).
    \label{entInt}
\end{align}
To second order in $\beta$, the first integral is
\begin{equation}
    \frac{t}{\rho_0} \int_{|v(\varepsilon)| t < \ell} d\varepsilon \varepsilon | \log \varepsilon |^3 \left ( \log 2 - \frac{\varepsilon^2 \beta ^2}{8} + \mathcal{O}(\beta^4) \right)
\end{equation}
which is much easier to deal with. To determine the integration bounds, we solve
\begin{equation}
    v(\varepsilon) t =\frac{\varepsilon |\log \varepsilon |^3 t}{\rho_0} = \ell
\end{equation}
for $\varepsilon>0$. Because $v(\varepsilon)$ does not increase monotonically with $\varepsilon$, there are multiple branches to the solution:
$
    \exp \left \lbrack 3 W_{-1}\left( \frac{-1}{3} \sqrt[3]{\frac{\ell}{t}} \right) \right \rbrack $, $
    \exp \left \lbrack 3 W\left( \frac{-1}{3} \sqrt[3]{\frac{\ell}{t}} \right) \right \rbrack$, and $
    \exp \left \lbrack 3 W\left( \frac{1}{3} \sqrt[3]{\frac{\ell}{t}} \right) \right \rbrack,
$
where $W_k(x)$ is the $k^{th}$ branch of the product log or Lambert W-function, and $W(x)$ is the principal branch of the product log function. The first two solutions are only valid (real) for $t>t^*= \frac{e^3 \ell}{27}$, while the third is valid for all $t>0$. Thus, $t^*$ is the time at which the slow modes begin to contribute to the entanglement growth. This set of slow modes makes the dynamics of the RSP markedly different than that of free fermions.

The second term in \eqref{entInt} can be integrated exactly. Taking into account, once again, the multiple domains of integration, and imposing an energy cutoff $\epsilon$ (which also functions as a velocity cutoff), we obtain a very complicated and unenlightening expression for the entanglement entropy, which we have used to fit the numerical data in Fig.~\ref{fig:RSPEEgrowth}.

% \bibliography{main}

\end{document}